\documentclass[journal=acs-an,manuscript=article]{achemso}

\usepackage[T1]{fontenc}

\usepackage{xcolor}
\definecolor{col_black}{RGB}{0,0,0}
\definecolor{col_review1}{RGB}{0,0,0} % {250,51,51} %  %\textcolor{review}{blabla...}
\definecolor{col_review2}{RGB}{0,0,0} %{51,51,250} %
\newcommand{\reviewfirst}[1]{\textcolor{col_review1}{#1}}

\usepackage{breakurl}
\usepackage{natbib}
\usepackage{url}
\author{Konstantinos Steiakakis}
\affiliation[tue]{Department of Mechanical Engineering, Eindhoven University of Technology, P. O. Box 513, 5600 MB Eindhoven, the Netherlands}

\author{Alan Pichard}
\affiliation[ur]{Univ. Rennes, CNRS, IPR - UMR 6251, Rennes, 35000, France}

\author{Maxime Vassaux}
\email{maxime.vassaux@univ-rennes.fr}
\affiliation[ur]{Univ. Rennes, CNRS, IPR - UMR 6251, Rennes, 35000, France}

\title{Molecular dynamics simulations reveal internal tension in native state collagen fibrils}
\begin{document}

\maketitle

\begin{abstract}
Collagen fibrils are the building block of many biological tissues, which viability depend on the fibrils properties. Altered properties of collagen fibrils are central to the appearance of many diseases, \reviewfirst{while} physiological or native properties must be reproduced for tissue engineering. Yet, the self-assembly, the structure, and therefore the properties of collagen fibrils remain elusive. One main reason is the extreme sensitivity of the fibrils to their environmental conditions, and in particular hydration which is only loosely bound by experimental measurements. Furthermore, mechanics are an integral part of the self-assembly process; \reviewfirst{forces exerted by cells or osmotic pressure} may result in internal stresses in collagen fibrils in native conditions. Here, we propose to investigate internal stresses in collagen fibrils by means of molecular dynamics simulations of the collagen microfibril model. Our simulations reveal the quantitative evolution of internal stresses in collagen fibrils with hydration. We establish a value of native hydration of collagen fibrils at $0.78$ g/g based on an absence of cross-sectional stresses. In turn, we determine a quantitative estimate of internal longitudinal stresses in collagen fibrils in native conditions of $210$ MPa. We find that internal longitudinal stresses are caused by an over-extended protein backbone rather than partial hydration, which appears remnant of the local forces driving collagen self-assembly. We also demonstrate the consequences of internal longitudinal stresses on the mechanical properties of collagen fibrils, which the absence of induces more than a $20\%$ decrease in the Young's modulus. Overall, our findings provide insights into the native structure and properties of collagen fibrils. More than ever, collagen fibrils appear to be assembled via an out-of-equilibrium process key to the synthesis of viable tissues.

\end{abstract}

%\tableofcontents

\section{Keywords}
collagen type I, internal stresses, molecular dynamics, self-assembly, hydration, fibrillar protein, native conditions

\section{Statement of significance}

The present study reveals the effect of hydration on the collagen fibril internal stresses and determines with unprecedented accuracy the hydration of collagen fibrils in their native state. With these findings, we revisit the current model of collagen fibrils self-assembly, accounting for internal stresses, and therefore the fact that collagen fibrils are out-of-equilibrium structures. Our conclusions provide a better understanding of the mechanical properties of collagen in native conditions and how these depend on hydration. These conclusions will improve our understanding of the physiological properties of fibrils, slight deviations in which are often the root of disease emergence. Further they will improve the viability of engineered tissues.

\section{Introduction}
\label{sec:intro}
Collagen fibrils are unique molecular structures synthesised by animals, found in many of their tissues such as tendons, ligaments, bones and cornea. \reviewfirst{Collagen fibrils are formed by the self-assembly of tropocollagens \cite{schmitt_new_1953}, known to be long and thin molecules with a distinctive triple helical structure made up of three alpha chains}. The internal structure of collagen fibrils has been largely documented from early work by North et al. \cite{north_structural_1954} and more recent reviews by Fratzl \cite{fratzl_collagen_2008}, and Shoulders and Raines \cite{shoulders_collagen_2009}. The structure of fibrils displays characteristic traits such as a discontinuous staggered longitudinal organization \cite{north_structural_1954} and and the so-called D-period \cite{smith_molecular_1968}. The rather disordered lateral organization of fibrils has been the subject of various packing \reviewfirst{hypotheses} \cite{hulmes_radial_1995}, which all suggest at short-range a quasi-hexagonal packing \cite{hulmes_quasi-hexagonal_1979}. The three-dimensional structure was only later resolved by Orgel et al. \cite{orgel_microfibrillar_2006} using X-ray fiber diffraction, supporting a microfibril crystalline subunit \cite{veis_limiting_1967}.

The self-assembly of collagen fibrils, or fibrillogenesis, is a complex process, inspiring for \reviewfirst{the design of biomimetic materials} \cite{shen_protein_2021}. Deficiencies in the self-assembly process, leading to altered 3D structure of collagen fibrils, are at the origin of several pathologies. Besides, in order to repair tissues or synthesise tissues in vitro it is necessary to control all the steps of the self-assembly process. In spite of great interest, the self-assembly process is still not entirely understood as many factors contribute: tropocollagen concentration, tropocollagen sequence (types of collagen involved), solvent composition (salt concentration, pH, proteoglycans concentration), and osmotic pressure. \reviewfirst{The influence of electrostatics has recently been evidenced using high-speed atomic force microscopy \cite{garcia-sacristan_operando_2024}.} Besides, fibrillogenesis may well be an out-of-equilibrium process \cite{bulavin_mechanism_2024}, therefore kinetics matter \cite{gisbert_high-speed_2021}. The large number of parameters controlling fibrillogenesis gives rise to a myriad of packing configurations leading to varying structures and therefore fibril mechanical properties \cite{jansen_role_2018, revell_collagen_2021}.

Another significant parameter of self-assembly is mechanical cues during self-assembly. The ability of cells, and in particular fibroblasts, to exert pulling forces has been shown to align fibrils \cite{canty_actin_2006}. Later, an in vitro study in absence of cells demonstrated that these pulling forces also align tropocollagen molecules to facilitate fibrillogenesis \cite{paten_flow-induced_2016}. Most recently, the capability of fibroblasts to exert pulling forces on tropocollagen has been observed in vitro \cite{silverman_tension_2024}. These studies confirm that fibrils are the result of an out-of-equilibrium self-assembly process, and are therefore expected to be out-of-equilibrium structures. As an illustration, residual strains estimated from variations of the D-spacing measured using X-ray diffraction have been documented in fibrils in the cartilage \cite{inamdar_secret_2017} and the bone-cartilage unit \cite{badar_nonlinear_2025}. This may suggest the occurrence of pulling forces and more largely mechanical cues during self-assembly may be the cause of residual strains or stresses in collagen fibrils that have yet to be documented and assessed.

In the present study, we aim to characterize potential residual stresses in collagen fibrils in native, \textit{i.e.}, physiological (\textit{in vivo}) conditions, in particular in native tissue hydration. Hydration in the native conditions is not precisely known for the fibril system, and it is known to influence internal stresses in collagen fibrils. Indeed compression in the longitudinal axis of the fibril arises from increased hydration \cite{masic_osmotic_2015}. Molecular dynamics simulations of the microfibril unit have proven to be a key tool over the last two decades to study collagen fibril structure and mechanics \cite{gautieri_hierarchical_2011, streeter_molecular_2011}, and more recently to study the specific effect of hydration \cite{vassaux_heterogeneous_2024}. Nonetheless, residual stresses have never been addressed. To that extent, we turn to large-scale molecular dynamics simulations of the microfibril structure determined from rat-tail tendon collagen fibrils in native conditions \cite{orgel_microfibrillar_2006}. Here, we specifically perform simulations at various hydration levels and assess associated internal stresses in the microfibril.

\section{Methods}
\label{sec:methods}

Our all-atom molecular model of the hydrated and neutralised collagen microfibril consists of the atoms constituting tropocollagen, counterions and water molecules. The position of the protein backbone atoms is set according to the microfibril crystallographic structure obtained by Orgel et al.\cite{orgel_microfibrillar_2006} using X-ray fiber diffraction on native rat tail tendon \textit{in situ}. The structure labelled \textit{3HR2} containing the position of the atoms can be found on the Protein Data Bank \cite{berman_protein_2000}. The molecular system is set up using periodic boundary conditions and the microfibril triclinic crystal unit\cite{orgel_microfibrillar_2006}. We modify amino acids from the \textit{3HR2} structure to match the exact sequence of human collagen type I as found in the single chains' sequences denoted \textit{NP\_000079} (for alpha-1(I) helices) and \textit{NP\_000080} (for the alpha-2(I) helices). We preserve hydroxyproline amino acids in the sequence of the chains according to the original structure. The ColBuilder server provides a straightforward interface to build such molecular models of collagen microfibrils \cite{obarska-kosinska_colbuilder_2021}. Then, we add $33$ negatively charged chlorine ions to neutralise the charge of the microfibril model. Last, we randomly pack water molecules inside the microfibril crystal unit to reach the desired hydration or weight of water to weight of protein ratio. The water molecules are inserted such that they are not less than $1.5$ \AA~away from any other atom in the volume to avoid inserting water molecules in between chains of the triple helix.

All-atom molecular dynamics (MD) simulations of the hydrated collagen microfibril are performed using the massively parallel LAMMPS software \cite{thompson_lammps_2022}. The potentials describing amino acid atoms interactions are modeled and parameterised with the CHARMM36m force field \cite{huang_charmm36m_2017} which has been extensively used for the simulation of protein structure. For verification purposes, we also model and parameterise interactions with the OPLS-AA force field \cite{robertson_improved_2015}. %The preparation of the parameter input files was facilitated by the use of the CHARMM-GUI server \cite{jo_charmm-gui_2008, lee_charmm-gui_2016}. 
Meanwhile, the potentials describing the interactions involving water atoms rely on the TIP3P model \cite{jorgensen_comparison_1983}. We use a rigid water model for computational efficiency, as we do not intend to focus on high-frequency bond and angle vibrational modes. Further, it has been widely used for protein simulations and recommended in combination with CHARMM36m \cite{boonstra_charmm_2016}. We perform ensemble MD simulations for better precision of the thermodynamic averages, the ensemble features $5$ replicas, each with different seeds to randomly initialise the velocities of the atoms and to randomly position the water molecules.

Equilibration simulations consist of NVT ensemble simulations at $308$ K. These simulations are performed for $20$ ns, which is shown to yield converged energies and stresses in the microfibril within $10$ ns (see figure \ref{fig:stress_hydration}.b). The simulation time remains rather short in comparison to typical biomolecular simulations because the initial structure of our simulations is derived from the crystalline structure of the protein \textit{in vivo}, and is therefore stable on microscopic timescales. Equilibration simulations are performed independently for all tested hydration levels. The equilibrated structures resulting from these simulations are employed as starting points for the following simulations of the perturbed microfibril model. Simulated perturbations include: ablation, ambient-pressure relaxation, and mechanical deformation. Ablation simulations consist of identical equilibration simulations except that the some covalent bonds are deleted in the core of the microfibril at the onset of the simulation. Covalent bonds are deleted between carbon and nitrogen atoms in the backbone of the tropocollagen, when bonded atoms are located at mid-span of the microfibril ($z \in [32.0, 32.5]$ nm). These NVT ensemble simulations are performed for $0.5$ ns until the motion of the protein stabilises. Ambient-pressure relaxation simulations consist of NPT ensemble simulations at $308$ K and $1$ bar. These simulations are performed for $5$ ns until the potential energy of the system converges and far beyond the complete elastic relaxation of internal stresses in the microfibril. Mechanical deformation simulations consist of out-of-equilibrium simulations at $308$ K, whereby the longitudinal $z$-dimension of the microfibril is increased at a constant strain rate throughout the simulation. We simulate two longitudinal strain amplitudes $\epsilon_{zz}$ ($0.05$, $0.20$) and two strain rates $\dot{\epsilon}_{zz}$ ($0.0025$, $0.01$ ns$^{-1}$). The lateral faces of the system are either left free or fixed during simulations.

The calculation of the global stress tensor relies on the Virial stress formula \cite{thompson_general_2009}. The stresses are sampled over the last $1$ ns during equilibration and relaxation simulations, while the stresses are sampled continuously during the mechanical deformation simulations. We focus on the diagonal components of the stress tensor, that is the stresses normal to the surfaces of the microfibril crystal unit. 

\section{Influence of hydration on internal stresses}

\begin{figure}
    \centering
    \includegraphics[width=\textwidth]{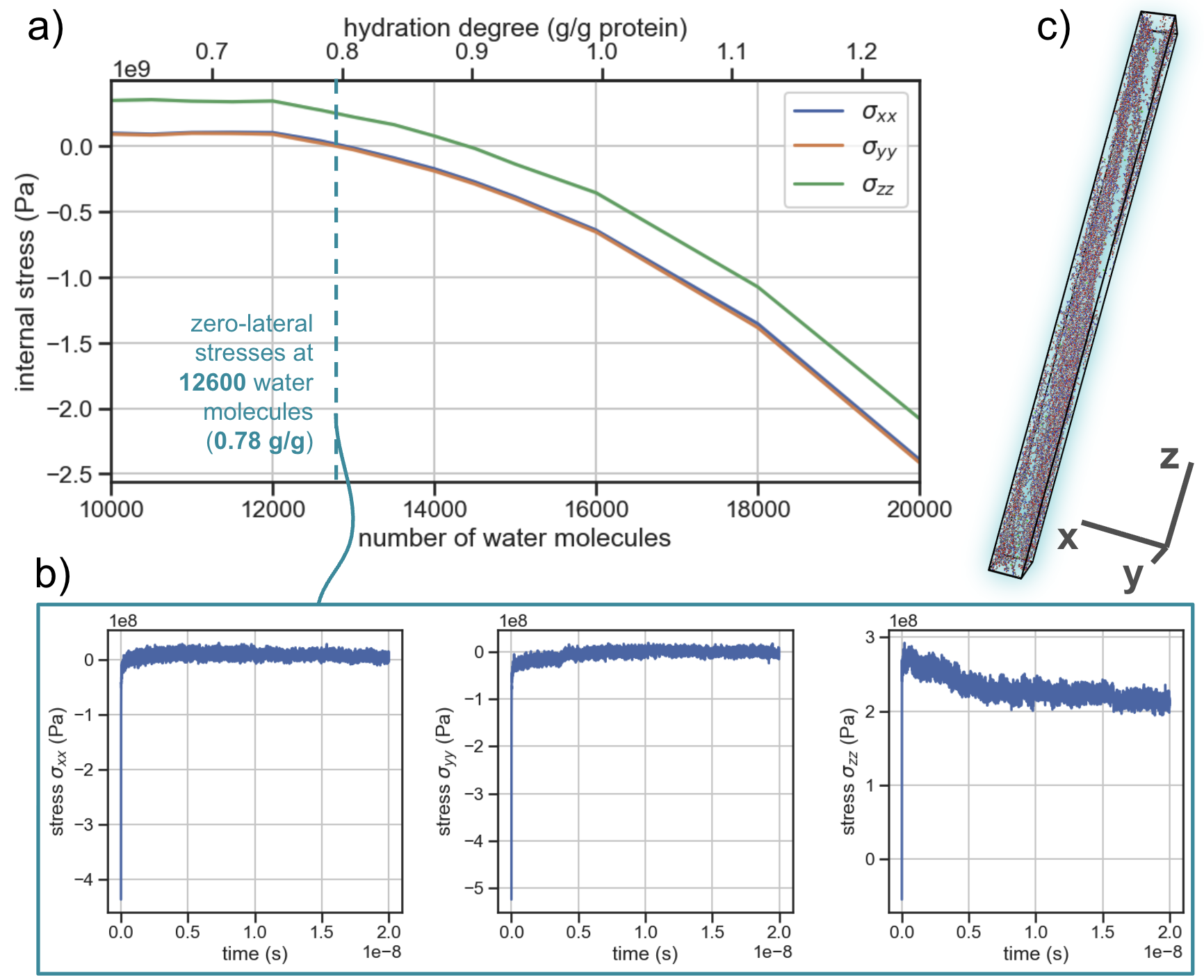}
    \caption{\textbf{Hydration and internal stresses.} (a) Evolution of the internal stresses in the microfibril sampled at $308$ K at varying hydration levels. The three diagonal components of the stress tensor $\sigma_{xx}$,  $\sigma_{yy}$, and $\sigma_{zz}$ are drawn in blue, orange and green, respectively. The internal stresses vary from tension to compression with increasing hydration. (c) Time-evolution of the internal stresses at $0.78$ g/g ($12600$ water molecules) up to equilibration, with lateral stresses converging to zero. At zero lateral stresses, the longitudinal stress is positive, that is the microfibril is under tension. (c) Visualisation of the structure of the microfibril at zero lateral stresses.}  
    \label{fig:stress_hydration}
\end{figure}

We now dive into the results of our molecular simulations and estimate the internal stresses in the collagen microfibril. We perform molecular dynamics simulations of the equilibration of the microfibril at hydration levels ranging from $0.6$ grams of water per gram of protein ($4000$ water molecules) to $1.25$ g/g ($20000$ water molecules). The mass of protein in the microfibril unit (an entire tropocollagen molecule of type I collagen) is $289.3$ kDa. We observe the evolution of the internal axial stresses normal to the $\underline{x}$, $\underline{y}$ and $\underline{z}$ directions, respectively $\sigma_{xx}$, $\sigma_{yy}$ and $\sigma_{zz}$. We perform equilibration simulations and observe the evolution of the stabilised stresses with hydration (see figure \ref{fig:stress_hydration}). First, the three diagonal components of the internal stress tensor all follow a similar nonlinear trend. Internal stresses are positive at low hydration, below $0.78$ g/g ($12600$ water molecules), implying internal tension in all three directions in the microfibril, and switch to compression with increasing hydration. The internal tension tends to saturate below $0.75$ g/g ($12000$ water molecules). Second, the microfibril is not transversely isotropic, nonetheless, the lateral stresses, that is $\sigma_{xx}$ and $\sigma_{yy}$, show identical values within error. The lateral stresses become null at an hydration level of about $0.78$ g/g ($12600$ water molecules). At this hydration level, the longitudinal stress $\sigma_{zz}$ is however far from null, about $210$ MPa, hinting at significant shrinking forces within the microfibril. Third, the longitudinal stress becomes null and negative at higher hydration, above $0.9$ \reviewfirst{g/g} ($14500$ water molecules). At such hydration, however, the lateral stresses are largely compressive. 

In order to verify our predictions of internal stresses, we perturb our model in several ways and compute the internal stresses again. We perform ensemble simulations whereby each replica of the ensemble consists of different random packing of water molecules and initial velocities (varying random seeds). Each replica in the ensemble yields almost identical internal stresses (see figure S2). Then, we compare the internal stresses calculations with values predicted using the OPLS-AA force-field. OPLS-AA yields some differences with the CHARMM36m force-field. Zero lateral pressure is reached at $12100$ water molecules, that is $500$ fewer molecules. Nonetheless, the converged longitudinal $\sigma_{zz}$ stress remains largely positive, about $170$ MPa. Last, we tested the influence of the temperature on the evolution and the stabilisation of internal stresses (see figure S3). We observe a decrease in the internal stresses when increasing the temperature from $290$ K to $330$ K. Although the longitudinal stress decays, it remains largely positive, above $200$ MPa. In short, all these verification simulations comfort the existence of shrinking forces in the microfibril at zero lateral pressure. This suggests that the existence of the shrinking does not arise as an artifact of the computational approach employed and is characteristic of the microfibril crystallographic structure obtained by Orgel et al.\cite{orgel_microfibrillar_2006}

\section{Native hydration and longitudinal stress}

Hydration is well-known to control the elastic properties of fibrils\cite{andriotis_collagen_2018}, their length when free to shrink, and their longitudinal stress when clamped \cite{masic_osmotic_2015}. In order to determine the longitudinal stress at native hydration, we need to determine native hydration first, that is the hydration of fibrils \textit{in vivo}. We here report hydration or water content estimations in grams of water per gram of protein or count of water molecules in the microfibril, which molar mass is $289270.4$ grams/mole. 

The earliest experimental estimations are based on wet to dry mass loss in tendon \cite{fullerton_orientation_1985}. The water content in the collagen fibril was estimated to be about $1.62$ g of water per gram of protein or $26034$ water molecules. The water content was measured by placing collagen samples in vacuum and heated at $90$°C until constant mass was observed. The study focused on tendon tissue, the lost water included not only water located in fibrils but also in between fibrils and fibres. Therefore, this value of hydration ($1.62$ g of water per gram of protein) would constitute an upper bound value. 

A more recent estimation can be based on variations of fibril cross-section measured using atomic force microscopy (AFM) \cite{andriotis_collagen_2018}. \reviewfirst{The measures compared native fibrils from mice tail tendon air-dried and fully hydrated in phosphate-buffered saline solution (PBS).} Assuming constant length of the fibrils during drying, we can compute a volume reduction from PBS to air of $0.473$ $\mu$m$^3$ per $\mu$m$^3$ of fibril. Using this ratio to compute the volume of water in the volume of the microfibril, we obtain $345.4$ nm$^3$ of water. The water content in the collagen fibril can therefore be estimated at $0.72$ gram of water per gram of protein or $11551.7$ water molecules. During drying in air, all water may not be removed; thus, this value of hydration ($0.72$ g of water per gram of protein) would constitute a lower bound value.

Combinations of the microfibril crystal unit measure from X-ray fiber diffraction \cite{orgel_microfibrillar_2006} and molecular dynamics simulations can also be used to estimate the native hydration. One possible criterion is to assume that native hydration corresponds to the amount of water required to stabilise the volume of the microfibril crystal unit in an isotropic ambient (1 atm) pressure barostat \cite{gautieri_hierarchical_2011}. This assumption leads to a water content in the collagen fibril estimated at $0.66$ gram of water per gram of protein or $10667$ water molecules. This hydration value deviates only marginally from the lower bound established from AFM measurements\cite{andriotis_collagen_2018}, but remains below this lower bound. We here propose an alternative criterion to determine native hydration using the same tools. A putative criterion could have been to have a stable volume in an anisotropic barostat, that is a barostat with dimensions of the crystal unit evolving independently. However, we previously showed that the three diagonal components of the internal stress tensor cannot be null simultaneously, at a given hydration level. Then, it becomes impossible to simulate a stable volume microfibril in an anisotropic barostat. We therefore propose to assume that native hydration corresponds to hydration leading to zero lateral stresses. This assumption leads to a water content in the collagen fibril estimated at $0.78$ gram of water per gram of protein or $12600$ water molecules. Quite strikingly, much like the previous simulated estimation, the hydration value deviates only about $1000$ water molecules per microfibril unit from AFM measurements on \textit{in vitro} fibrils and remains, this time, above the experimental lower bound.

Concluding on the estimation of the internal longitudinal stress at native hydration, AFM measurements, stable volume assumption and zero lateral stress assumption would yield values about $330$ MPa, $350$ MPa and $210$ MPa, respectively.

\section{Looking for the origin of the longitudinal stress}

At this stage, the internal longitudinal stress at native hydration can be imparted to either potential voids in the water phase, collapsing tropocollagen molecules caused by entropic forces, or over-stretched covalent bonds. It appears impossible to find hydration levels for which the three diagonal components of the internal stress tensors are null, that is an entirely stable collagen microfibril. We now attempt to discern the origin of the significant tensile longitudinal stress within the microfibril at the supposed native hydration, or at least at zero lateral pressure. Following the original hypothesis that collagen fibrils are assembled under stretch, we aim to determine whether the covalent bonds in the backbone of the tropocollagens are overly stretched. To that extent, we perform two separate tests aiming at relaxing the tension inside the tropocollagen, and then we measure the structural and thermodynamic consequences of such relaxation.

\subsection{Protein-ablation stress relaxation}
\begin{figure}
    \centering
    \includegraphics[width=\textwidth]{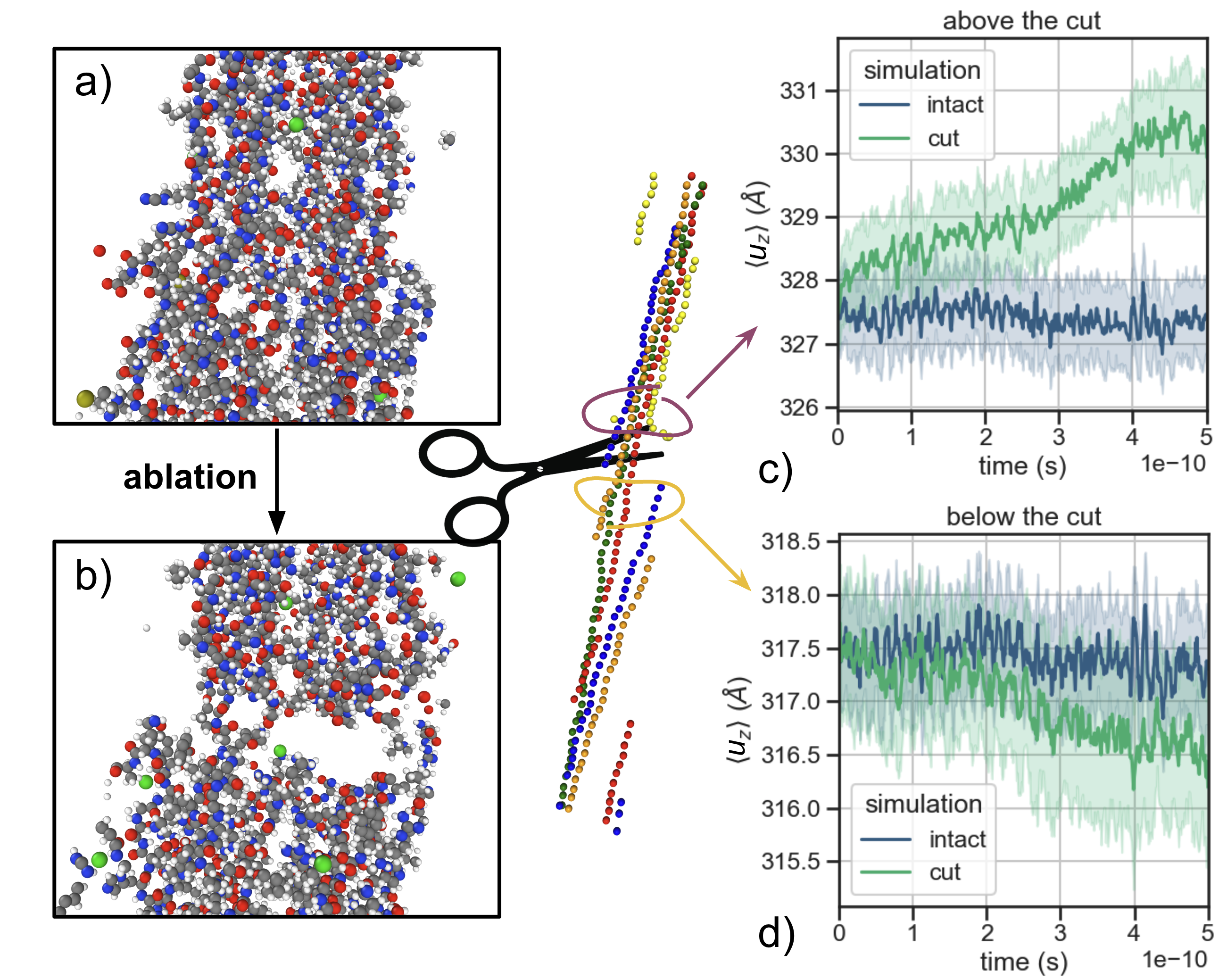}
    \caption{\textbf{Ablation of the microfibril.} Influence of the removal of covalent bonds in the tropocollagen molecules in the whole cross-section for $z$ ranging from $32$ to $32.5$ nm. Visualisations of the microfibril in the region of the cut, (a) before and (b) after relaxation of the microfibril. Evolution of the average $z$ position $\langle u_z \rangle$ of the atoms initially located (c) above the cut ($z$ ranging from $32.5$ to $33$ nm) and (d) below the cut ($z$ ranging from $31.5$ to $32$ nm). The evolution of $\langle u_z \rangle$ is compared in presence ("cut") and in absence ("intact") of the removal of covalent bonds.}
    \label{fig:tropo_cut}
\end{figure}

We perform a simulation of the dynamics of the microfibril subject to a cut in the region where the microfibril is constituted only of four tropocollagen strands (see figure \ref{fig:tropo_cut}). We observe the average longitudinal position $\langle u_z \rangle$ of the protein atoms just above (atoms with initial longitudinal positions ranging from $32.5$ nm to $33$ nm) and below (atoms with initial longitudinal positions ranging from $31.5$ nm to $32$ nm) the cut (see figure \ref{fig:tropo_cut}.c and \ref{fig:tropo_cut}.d, respectively). We compare these averaged displacements (green curve) with the averaged displacements of the same atoms in absence of the cut (blue curve). In absence of the cut, the position of the atoms remains stable and fluctuates, above the cut around $32.75$ nm and below the cut around $31.75$ nm, throughout the whole simulation. In contrast, in presence of the cut, the position of the atoms rapidly drift away from the region of the cut. Considering the mean displacement of atoms above and below the cut during the span of the simulation ($0.3$ nm and $0.1$ nm, respectively), we can estimate an average drifting speed of $0.8$ m/s.

The rather rapid motion of the atoms in the vicinity of the cut tends to support a recoil of the tropocollagen strands in the microfibril associated with a release of elastic energy. In other words, the backbone of the protein triple helices as found in vivo, during X-ray fiber diffraction imaging, is out of equilibrium, either stretched, bent or twisted.

% # Can we get data at longer timescales? Sample for longer, maybe 1 ns?

\subsection{Ambient-pressure stress relaxation}

% edit subsection with updated figure including local stresses

\begin{figure}
    \centering
    \includegraphics[width=\textwidth]{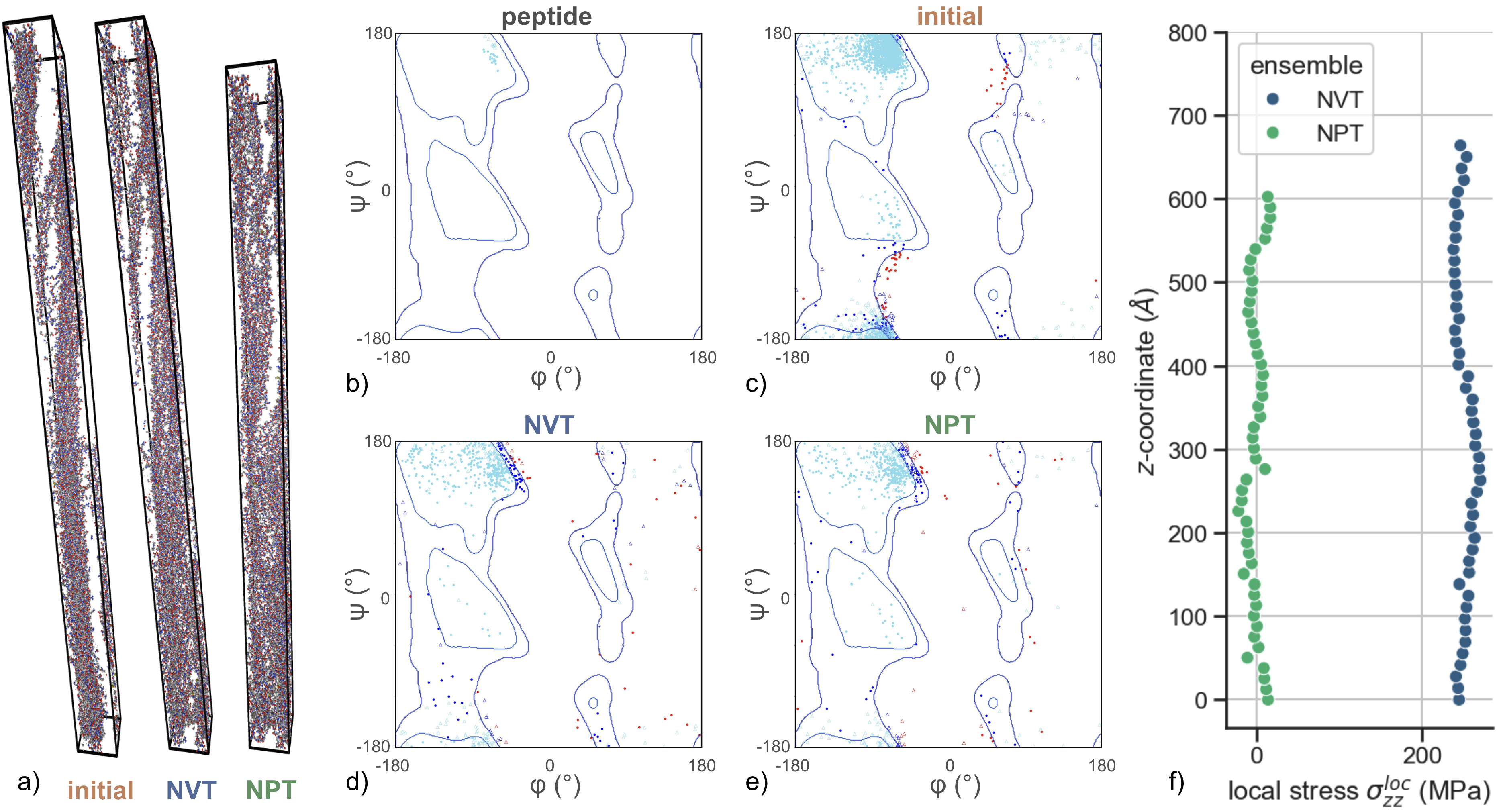}
    \caption{\textbf{Release the stresses.} Comparison of the structures of the microfibril from X-ray fiber diffraction (initial), after equilibration (NVT-ensemble) and after ambient-pressure stress relaxation (NPT-ensemble). (a) Visualisation of the atomistic structure for each of the three states of the microfibril (initial, NVT equilibration, NPT relaxation). Ramachandran plots of the torsion (dihedral) angles ($\phi$ and $\psi$) of the amino acids in tropocollagen molecules. The distributions of the torsion angles are computed for (b) a short 30 amino acid long peptide of collagen (PDB: 7CWK) and the (c) initial, (d) equilibrated and (e) relaxed structures of the microfibril. Cyan, dark blue, and red dots represent torsion angles of favoured, allowed and disallowed regions respectively. Contour lines delimit the different regions based on energy calculations. Dots represent residues other than glycine, while triangles indicate glycine residues.
    (f) Evolution of the local stress $\sigma^{loc}_{zz}$ in the protein only along the longitudinal $z$-dimension in the microfibril before (NVT) and after (NPT) ambient-pressure relaxation.}
    \label{fig:relaxation}
\end{figure}

We compare three different states of the microfibril: (i) the initial state, as imaged by X-ray fiber diffraction \cite{orgel_microfibrillar_2006}, (ii) the equilibrated state, resulting from a NVT-ensemble simulation, and (iii) the relaxed state, reached after a NPT-ensemble at ambient pressure simulation. For each state, we visualise the structure of the microfibril and we compute the Ramachandran plots as well as local stresses along the longitudinal dimension (see figure \ref{fig:relaxation}).

The evolution of the atomistic structure of the microfibril (see figure \ref{fig:relaxation}.a) reveals little change induced by the equilibration. In drastic contrast, we observe a collapse of the microfibril structure when we remove the constant-volume constraint. The simulation of ambient pressure stress relaxation leaves all dimensions of the microfibril free to change. As a result of the internal tension in the longitudinal direction, the microfibril shrinks. %Inversely, the detailed energy contributions reveal rather significant changes (see figure \ref{fig:relaxation}.b,c) during the equilibration. Nonetheless, we do observe a decrease of the total and potential energies, the later decreases from $E^{NVT}_{pot} = -105086.3$ to $E^{NPT}_{pot} = -105952.9$ kcal/mole. This reduction is the result of the decrease of the angle and dihedral energies, the bond energy increasing slightly.

A potential explanation for the internal tension in the microfibril consists of a distorted protein backbone; in turn we revert to Ramachandran plots (see figure \ref{fig:relaxation}.b,c,d,e). Ramachandran plots are computed using RamPlot \cite{kumar_ramplot_2025}. The plots are two-dimensional distributions of the two torsion angles found in the backbone of each amino acid of a protein. Samples in the Ramachandran plots are classified in three categories (favored, unfavored and disallowed) according to their probability of existence, inversely proportional to their potential energy. Ramachandran plots are used as a complementary local measure of structural changes, which can be assessed globally qualitatively with structure visualisation and quantitatively with potential energy calculations. The Ramachandran plot for a short triple helix (PDB:7CWK), also called a collagen mimetic peptide, is provided as a reference distribution of torsion angles (see figure \ref{fig:relaxation}.b). All samples from the distribution are located in the specific part of the top left quadrant associated with polyproline-II helices. The plot for the initial structure obtained from X-ray fiber diffraction (PDB:3HR2) \cite{orgel_microfibrillar_2006} displays a much more scattered distribution of torsion angles (see figure \ref{fig:relaxation}.c). The distribution contains now about 3000 pairs of dihedral angles, most of them concentrated in the polyproline-II specific region, but also in the center of the left half of the plot, specific to alpha helices. One significant group of samples is located in between the polyproline-II and the alpha helices specific regions. Another significant group of samples is located in the top right quadrant, in between and around the regions specific of left-handed alpha helices and left polyproline-II conformations. Both groups of samples reveal amino acids with disallowed configuration (red) of backbone torsion, that is with high potential energy. The plot for the equilibrated structure (see figure \ref{fig:relaxation}.d) displays a few differences with that of the initial structure. The mostly favored (cyan) configurations associated with polyproline-II helices have now spread, displaying unfavored (dark blue) and disallowed configurations with lower $\phi$ angle values. The density of samples with disallowed configurations in the bottom left and the top right is drastically reduced. In the region between the polyproline-II and alpha helix regions, the amino acids with disallowed configurations appear to have transitioned to unfavored configurations. \reviewfirst{The evolution of samples} with disallowed configurations in the top right quadrant is less obvious, yet their reduction correlates with the apparition of scattered samples with disallowed configurations throughout the right half of the plot. The plot for the relaxed structure (see figure \ref{fig:relaxation}.e) reveals more subtle changes in the configuration of amino acids. Most of the amino acids with unfavored configurations trapped between polyproline-II and alpha helix configurations disappear, transitioning to either one of the two configurations. \reviewfirst{Similarly} for most of the disallowed configurations in the right half of the plot. These configurational or conformational changes have the main consequence of lowering the potential energy of the tropocollagen molecule. \reviewfirst{In short, the initial structure displayed a highly scattered distribution of torsion angles, with significant clusters in polyproline-II and alpha-helix regions, as well as disallowed conformations in transitional and left-handed regions. After equilibration, disallowed conformations decreased, especially between polyproline-II and alpha-helix regions, while some shifted to unfavored states, and the density of disallowed samples in the top right and bottom left quadrants dropped. Last, in the relaxed structure, most unfavored and disallowed conformations transition to favored states, further lowering the molecule’s potential energy.}

The detailed analysis of the Ramachandran plots and local stresses reinforces the hypothesis that tension is associated with an out-of-equilibrium protein in the microfibril, rather than partial hydration (voids). The idealised structure of tropocollagen, the triple helix, should only display amino acids in the region at $(\phi = -75^{\circ}, \psi = +145^{\circ})$ as seen from the short peptide analysis. The tropocollagen in the initial structure deviates from this idealised configuration. Nonetheless, the subsequent equilibration (NVT-ensemble simulation) and relaxation (NPT-ensemble simulation) help to recover the idealised structure, with subtle dihedral angle variations, progressively reducing the number of amino acids in other quadrants of the Ramachandran plot. It has to be noted, that although these angle variations are subtle, they yield significant stress relaxation. 

In order to fully confirm the hypothesis of a stretched protein backbone, we last focus on local estimations of the longitudinal stress in the protein along the longitudinal dimension of the microfibril (see figure \ref{fig:relaxation}.f). The local stress tensor, as opposed to the global stress tensor discussed so far, is estimated averaging the per-atom stress tensor of the protein atoms over time and space \cite{thijssen_necking_2023}. The local stresses provide an alternative measure of the tension to global stresses measured so far. The local stresses allow to check whether localities that would induce stresses, for example cavitation, may exist. Here, we average over $x$-$y$ cross-sections of one fiftieth of the $z$-dimension of the microfibril, for $0.25$ ns. We compare the distributions of the local stress $\sigma^{loc}_{zz}$ for the NVT-equilibrated and the NPT-relaxed structures. We observe the entire relaxation of the local stresses in the protein after ambient-pressure relaxation. In the NVT-equilibrated system, we also note that the average local stress in the protein averages to $210$ MPa, which is almost identical to the global stress $\sigma_{zz}$ (see figure \ref{fig:stress_hydration}.b, right). This further confirms that the internal longitudinal stress in the microfibril is caused by the stretching of the protein backbone rather than partial hydration.

% Is the main change between NVT and NPT associated with change in entropy? Look at kinetic energy changes... Is it relevant to look at kinetic energy as there does not seem to be any change in kinetic energy? How can we probe conformational entropy in MD simulations?
% Check the why there seem to be less samples in the NVT and NPT plots.
% Can we say there are more alpha configurations than PP? (which are shorter for the same amount of AA)

\section{Mechanical properties in presence of longitudinal stresses}

\begin{figure}
    \centering
    \includegraphics[width=\textwidth]{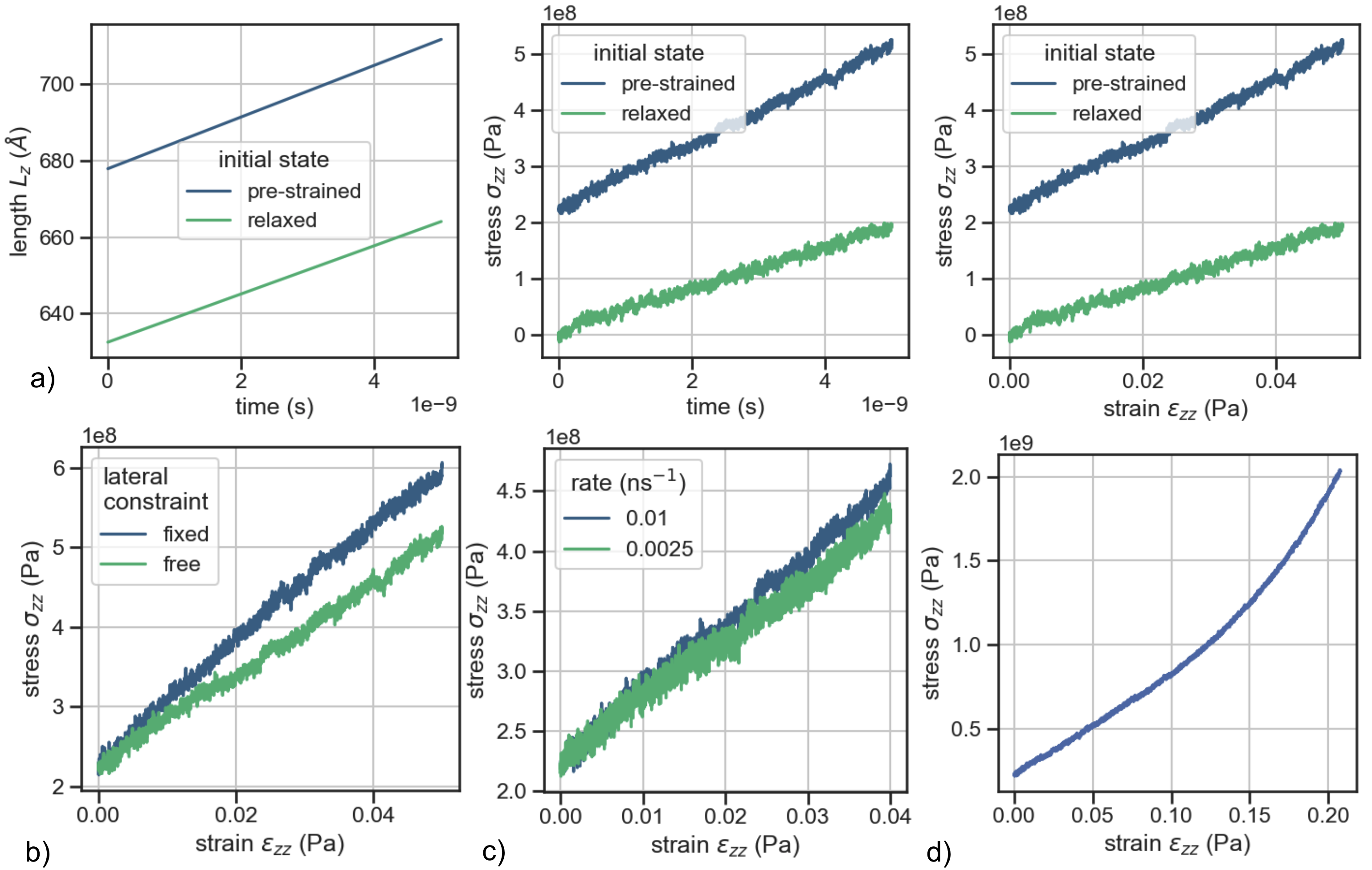}
    \caption{\textbf{Deformation of the microfibril.} Simulations of constant rate deformation applied to the microfibril in the longitudinal $z$ direction. (a) Comparison between the relaxed and the internally stressed initial state for (left) the evolution of the length $L_z$ of the microfibril with time; (middle) the evolution of the longitudinal stress $\sigma_{zz}$ with time; (right) evolution of the stress $\sigma_{zz}$ with the strain $\epsilon_{zz}$. Influence on the longitudinal stress $\sigma_{zz}$ of (b) the presence of lateral constraints (free vs. fixed), (c) the strain rate ($\dot{\epsilon_{zz}} = 0.01$ ns$^{-1}$ vs. $\dot{\epsilon_{zz}} = 0.0025$ ns$^{-1}$), or (d) an extended strain amplitude up to $20$ \%.} 
    \label{fig:mechanics}
\end{figure}

\reviewfirst{We now move away} from the origin of the internal longitudinal stress in the microfibril to explore the consequences of its existence on the mechanical properties and the response to longitudinal stretch of collagen fibrils (see figure \ref{fig:mechanics}). \reviewfirst{The deformation or strain along the z-axis $\epsilon_{zz}$ is defined as the ratio between the length variation and the initial length of the microfibril.} The deformation $\epsilon_{zz}$ is applied at constant rate on the microfibril in two different initial states: internally stressed (pre-stressed) and relaxed. These two states correspond to the two final states resulting from the NVT and NPT simulations, respectively, as shown in figure \ref{fig:relaxation}.a (right, left). The lateral boundaries of the microfibril are free to move and the deformation rate is $\dot{\epsilon_{zz}} = 0.01$ ns$^{-1}$. As expected, the relaxed microfibril starts from a shorter length about $63.5$ nm (green curve on figure \ref{fig:mechanics}.a), than the microfibril we denote as pre-stressed ($67.7$ nm, blue curve on figure \ref{fig:mechanics}.a). Besides, the complete relaxation of the microfibril denoted as relaxed is confirmed as the initial stress $\sigma_{zz}$ is null. When the constant rate deformation $\epsilon_{zz}$ is applied, both cases display linear elastic responses. \reviewfirst{We compute the Young's modulus of the collagen fibril as the average slope of the stress-strain curve between $\epsilon_{zz}=0$ and $\epsilon_{zz}=0.05$.} The Young's modulus of the collagen fibril is computed at $5.6$ GPa when the fibril presents internal longitudinal stresses (pre-stressed), while the Young's modulus decreases down to $3.8$ GPa when the fibril is relaxed prior to applying the deformation. Our results show a Young's modulus an order of magnitude higher for a hydrated microfibril than Gautieri et al. \cite{gautieri_hierarchical_2011}, who predicted a modulus in agreement with measurements from cross-sectional AFM \cite{van_der_rijt_micromechanical_2006} and half the value ($0.86$ GPa) found using microelectromechanical systems\cite{shen_stress-strain_2008}. Our results agree well with Young's modulus calculations of single tropocollagen molecules from Brillouin light scattering \cite{cusack_determination_1979}, X-ray diffraction measurements \cite{sasaki_stress-strain_1996} as well as steered \cite{lorenzo_elastic_2005} or coarse-grained \cite{gautieri_coarse-grained_2010} MD simulations. However, these measurements do not allow the mechanism of deformation associated with sliding of tropocollagens past each other.

We further investigate the parameters of the mechanical deformation simulations on the internally stressed microfibril. We compare the influence of the lateral boundary conditions on the stress-strain response. We apply free (green) and fixed (blue) lateral conditions and observe a quarter smaller final stress in presence of free lateral microfibril boundaries (see figure \ref{fig:mechanics}.b). In turn, the fixed lateral boundary conditions yield a stiffness of $7.2$ GPa (and $5.6$ GPa in the free boundaries case). This comparison reveals that the lower average stiffness in the case of free lateral surfaces is associated with a transient change of slope around $\epsilon_{zz} = 0.02$. This may be associated with sudden structural changes in the protein or in the water, in particular disruption of the hydrogen bond network. Since the slope appears to recover its original value after the change of slope at  $\epsilon_{zz} = 0.02$, we may be facing a stick-slip mechanism between tropocollagen strands. As already discussed, our stiffness values, in particular under free lateral surfaces (Young's modulus) are high and correspond to values of stretching tests performed on single tropocollagen molecules, that is in absence of intermolecular sliding. Nonetheless, we do seem to observe intermolecular sliding in the free lateral surfaces scenario at $\epsilon_{zz} = 0.02$.

We also investigate the influence of the rate of deformation $\dot{\epsilon_{zz}}$ (see figure \ref{fig:mechanics}.c). We slow the rate of deformation four times down to $\dot{\epsilon_{zz}} = 0.0025$ ns$^{-1}$. The average stiffness measure over the course of the applied \reviewfirst{strain} only slightly decreases when the lower \reviewfirst{strain} rate is applied, from $5.6$ to $5.2$ GPa. Further, we observe that the transient change of slope around $\epsilon_{zz} = 0.02$ is accentuated when reducing the rate. Although, the effect of the deformation rate on the deformation and the stiffness is significant, it is less so than the effect of the initial internal longitudinal stress. We do not expect that much slower \reviewfirst{strain} rates affordable within microsecond-long simulations would yield Young's moduli comparable with AFM and MEMS measurements.

Last, we extended the amplitude of the applied \reviewfirst{strain} in the reference case with internal pre-stress (see figure \ref{fig:mechanics}.d). Beyond $0.1$ \reviewfirst{strain}, the microfibril displays the expected non-linear hyperelastic behaviour as documented by Gautieri et al.\cite{gautieri_hierarchical_2011}, associated with an increase of the instantaneous modulus of the material. Nonetheless, up to $0.2$ we do not observe softening of the material. Since rupture of covalent bonds is not allowed in such an empirical force-field, failure of single amino acid chains was not expected to occur, yet we neither observe denaturation of the tropocollagen molecules nor significant sliding of tropocollagens past each other.

In brief, the internal pre-stress may have several positive effects on the mechanical properties of the collagen fibrils. The internal pre-stresses tend to increase the initial alignment of the tropocollagens. The straightened tropocollagens would also have drastically increased initial stiffness, as elastic energy is either used to stretch peptidic covalent bonds or to increase steric interactions between single helices. Moreover, the internal pre-stresses, contributing to the alignment of tropocollagen molecules, increase their surface of contact and therefore enhance frictional interactions much like what can be observed in yarn \cite{seguin_twist-controlled_2022}.

% - Influence of the internal tension on fatigue (variation of Young's modulus), on resonance frequency (damping of mechanical sollicitation).
%   - residual strains in collagen after loading: Andriotis/ActaBiomat

% # What is the evolution of the lateral dimensions during pulling with free lateral surfaces?
% # Should we cut the tropocollagen at the crossing of the PBC ?

\section{Conclusions}
\label{sec:conclusion}

\reviewfirst{The purpose of the present study was to explore the presence of internal stresses in collagen fibrils resulting from native self-assembly using molecular dynamics simulations.} As the physiological or native hydration of collagen fibrils \textit{in vivo} is not quantitatively established, we studied internal stresses for various hydration levels. For this purpose, we built a molecular model of the collagen type I microfibril, of which we simulated the hydration and equilibration at hydration levels ranging from $0.6$ to $1.25$ gram of water per gram of protein. 

We found that the microfibril structure cannot be found devoid of internal stresses. Either showing compressive lateral stresses at zero longitudinal stress or tensile longitudinal stress at zero lateral stresses. Overall, internal stresses were found to shift from tension to compression with increasing hydration levels. We established the quantification of native hydration from different techniques, that is previous experimental and computational works, as well as from our own methodology. We found values for native hydration in the range of $0.60$ to $0.80$ g/g, and we proposed a criterion based on zero lateral stresses in the collagen fibril to establish native hydration, yielding a value of $0.78$ g/g. In turn, we were able to determine the value of the longitudinal internal stress in the collagen fibril at native hydration of about $210$ MPa.

We further investigated the origin of internal longitudinal stress in the collagen fibril simulating their relaxation via protein ablation or ambient-pressure control. We found from these out-of-equilibrium simulations that the internal stresses rather result from an overstretched protein backbone (distorted polyproline-II helices) and entropic effects rather than capillary effects associated with partial hydration. We finalised this study computing the effect of longitudinal stresses on the longitudinal Young's modulus of the hydrated microfibril. We found that the release of internal stresses yielded a decrease of $22$\% of the modulus as well as a decrease of $7$\% of the microfibril length.

Our study challenges the validity of past molecular models of collagen fibrils, which did not consider the existence of internal stresses and employed artificial constraints to ensure the fibril stability. And most importantly, our study brings novel fundamental knowledge on collagen fibrils, in particular on their structural and mechanical properties \textit{in vivo}. The existence of internal longitudinal stresses at native hydration comforts that the collagen fibrils are out-of-equilibrium structures self-assembled through complex mechanisms involving mechanical forces. Such a finding is key for the design of improved biomimetic collagenous materials and tissue engineering applications and may explain how large fibrils are assembled for anisotropic tissues such as tendons.

\section{Appendices}

The supporting information contains supplementary figures showing the detail evolution of internal stresses (i) in an ensemble of replicas; (ii) with the OPLS-AA force-field; and (iii) with temperature. The data serves as a verification of the molecular model.

\section{Data and code availability}

A GitHub repository containing data and scripts has been created (\url{https://github.com/mvassaux/collagen_tension/}). The data includes structures of the microfibril equilibrated at varying water contents as well as the temporal evolution of the thermodynamic variables in these simulations. The scripts feature the workflow to hydrate, equilibrate (NVT), relax (NPT) and deform a molecular model of the microfibril. The scripts also feature the data analysis pipeline implemented in a Python notebook.

\section{Authors contributions}

% K.S: Modelling, Editing; A.P: Simulations; M.V: Conceptualization, Modelling, Simulations, Writing, Editing.
K.S: Methodology, Validation, Writing – review \& editing; A.P: Software, Investigation; M.V: Project administration, Conceptualization, Resources, Supervision, Methodology, Software, Investigation, Data curation, Formal analysis, Validation, Visualization, Writing – original draft, Writing – review \& editing, Funding acquisition.

\section{Declaration of interests}

The authors declare no competing interests.

\section{Acknowledgements}
\label{sec:acknowledgements}

The authors acknowledge funding support from the Fondation ARC through its postdoctoral fellowship program (grant number ARCPDF22020010001195). The authors also acknowledge funding support from CNRS through its Tremplin program. This project was provided with HPC and storage resources by GENCI at TGCC on the supercomputer Joliot Curie's SKL partition (grant number 2024-13939).

\newpage

\bibliography{references.bib}

@article{gautieri_hierarchical_2011,
	title = {Hierarchical {Structure} and {Nanomechanics} of {Collagen} {Microfibrils} from the {Atomistic} {Scale} {Up}},
	volume = {11},
	issn = {1530-6984},
	url = {https://doi.org/10.1021/nl103943u},
	doi = {10.1021/nl103943u},
	number = {2},
	urldate = {2022-03-17},
	journal = {Nano Letters},
	author = {Gautieri, Alfonso and Vesentini, Simone and Redaelli, Alberto and Buehler, Markus J.},
	month = feb,
	year = {2011},
	pages = {757--766},
}

@article{streeter_molecular_2011,
	title = {A molecular dynamics study of the interprotein interactions in collagen fibrils},
	volume = {7},
	url = {https://pubs.rsc.org/en/content/articlelanding/2011/sm/c0sm01192d},
	doi = {10.1039/C0SM01192D},
	language = {en},
	number = {7},
	urldate = {2022-03-17},
	journal = {Soft Matter},
	author = {Streeter, Ian and Leeuw, Nora H. de},
	year = {2011},
	pages = {3373--3382},
}

@article{seguin_twist-controlled_2022,
	title = {Twist-{Controlled} {Force} {Amplification} and {Spinning} {Tension} {Transition} in {Yarn}},
	volume = {128},
	url = {https://link.aps.org/doi/10.1103/PhysRevLett.128.078002},
	doi = {10.1103/PhysRevLett.128.078002},
	number = {7},
	urldate = {2022-04-11},
	journal = {Physical Review Letters},
	author = {Seguin, Antoine and Crassous, Jérôme},
	month = feb,
	year = {2022},
	pages = {078002},
}

@article{andriotis_collagen_2018,
	title = {Collagen {Fibrils}: {Nature}’s {Highly} {Tunable} {Nonlinear} {Springs}},
	volume = {12},
	issn = {1936-0851},
	shorttitle = {Collagen {Fibrils}},
	url = {https://doi.org/10.1021/acsnano.8b00837},
	doi = {10.1021/acsnano.8b00837},
	number = {4},
	urldate = {2022-11-09},
	journal = {ACS Nano},
	author = {Andriotis, Orestis G. and Desissaire, Sylvia and Thurner, Philipp J.},
	month = apr,
	year = {2018},
	pages = {3671--3680},
}

@article{jorgensen_comparison_1983,
	title = {Comparison of simple potential functions for simulating liquid water},
	volume = {79},
	issn = {0021-9606},
	url = {https://doi.org/10.1063/1.445869},
	doi = {10.1063/1.445869},
	number = {2},
	urldate = {2023-06-06},
	journal = {The Journal of Chemical Physics},
	author = {Jorgensen, William L. and Chandrasekhar, Jayaraman and Madura, Jeffry D. and Impey, Roger W. and Klein, Michael L.},
	month = jul,
	year = {1983},
	pages = {926--935},
}

@article{berman_protein_2000,
	title = {The {Protein} {Data} {Bank}},
	volume = {28},
	issn = {0305-1048},
	url = {https://doi.org/10.1093/nar/28.1.235},
	doi = {10.1093/nar/28.1.235},
	number = {1},
	urldate = {2023-06-06},
	journal = {Nucleic Acids Research},
	author = {Berman, Helen M. and Westbrook, John and Feng, Zukang and Gilliland, Gary and Bhat, T. N. and Weissig, Helge and Shindyalov, Ilya N. and Bourne, Philip E.},
	month = jan,
	year = {2000},
	pages = {235--242},
}

@article{obarska-kosinska_colbuilder_2021,
	title = {{ColBuilder}: {A} server to build collagen fibril models},
	volume = {120},
	issn = {0006-3495},
	shorttitle = {{ColBuilder}},
	url = {https://www.cell.com/biophysj/abstract/S0006-3495(21)00561-0},
	doi = {10.1016/j.bpj.2021.07.009},
	language = {English},
	number = {17},
	urldate = {2023-11-19},
	journal = {Biophysical Journal},
	author = {Obarska-Kosinska, Agnieszka and Rennekamp, Benedikt and Ünal, Aysecan and Gräter, Frauke},
	month = sep,
	year = {2021},
	pmid = {34265261},
	pages = {3544--3549},
}

@article{gautieri_coarse-grained_2010,
	title = {Coarse-{Grained} {Model} of {Collagen} {Molecules} {Using} an {Extended} {MARTINI} {Force} {Field}},
	volume = {6},
	issn = {1549-9618},
	url = {https://doi.org/10.1021/ct100015v},
	doi = {10.1021/ct100015v},
	number = {4},
	urldate = {2024-03-11},
	journal = {Journal of Chemical Theory and Computation},
	author = {Gautieri, Alfonso and Russo, Antonio and Vesentini, Simone and Redaelli, Alberto and Buehler, Markus J.},
	month = apr,
	year = {2010},
	pages = {1210--1218},
}

@incollection{fratzl_collagen_2008,
	address = {Boston, MA},
	title = {Collagen: {Structure} and {Mechanics}, an {Introduction}},
	isbn = {978-0-387-73906-9},
	shorttitle = {Collagen},
	url = {https://doi.org/10.1007/978-0-387-73906-9_1},
	language = {en},
	urldate = {2024-04-02},
	booktitle = {Collagen: {Structure} and {Mechanics}},
	publisher = {Springer US},
	author = {Fratzl, P.},
	editor = {Fratzl, Peter},
	year = {2008},
	doi = {10.1007/978-0-387-73906-9_1},
	pages = {1--13},
}

@article{fullerton_orientation_1985,
	title = {Orientation of tendons in the magnetic field and its effect on {T2} relaxation times.},
	volume = {155},
	issn = {0033-8419},
	url = {https://pubs.rsna.org/doi/abs/10.1148/radiology.155.2.3983395},
	doi = {10.1148/radiology.155.2.3983395},
	number = {2},
	urldate = {2024-04-15},
	journal = {Radiology},
	author = {Fullerton, G D and Cameron, I L and Ord, V A},
	month = may,
	year = {1985},
	pages = {433--435},
}

@article{masic_osmotic_2015,
	title = {Osmotic pressure induced tensile forces in tendon collagen},
	volume = {6},
	copyright = {2015 The Author(s)},
	issn = {2041-1723},
	url = {https://www.nature.com/articles/ncomms6942},
	doi = {10.1038/ncomms6942},
	language = {en},
	number = {1},
	urldate = {2024-05-15},
	journal = {Nature Communications},
	author = {Masic, Admir and Bertinetti, Luca and Schuetz, Roman and Chang, Shu-Wei and Metzger, Till Hartmut and Buehler, Markus J. and Fratzl, Peter},
	month = jan,
	year = {2015},
	pages = {5942},
}

@article{bulavin_mechanism_2024,
	title = {Mechanism of protofibril formation in aqueous collagen solutions},
	volume = {14},
	issn = {2158-3226},
	url = {https://doi.org/10.1063/5.0238555},
	doi = {10.1063/5.0238555},
	number = {11},
	urldate = {2024-11-18},
	journal = {AIP Advances},
	author = {Bulavin, Leonid A. and Cherevko, Kostyantyn V. and Khorolskyi, Oleksii V. and Svechnikova, Oksana S. and Zabashta, Yurii F.},
	month = nov,
	year = {2024},
	pages = {115116},
}

@article{silverman_tension_2024,
	title = {Tension in the ranks: {Cooperative} cell contractions drive force-dependent collagen assembly in human fibroblast culture},
	volume = {7},
	issn = {2590-2385},
	shorttitle = {Tension in the ranks},
	url = {https://www.sciencedirect.com/science/article/pii/S2590238524000237},
	doi = {10.1016/j.matt.2024.01.023},
	number = {4},
	urldate = {2024-11-21},
	journal = {Matter},
	author = {Silverman, Alexandra A. and Olszewski, Jason D. and Siadat, Seyed Mohammad and Ruberti, Jeffrey W.},
	month = apr,
	year = {2024},
	pages = {1533--1557},
}

@article{jansen_role_2018,
	title = {The {Role} of {Network} {Architecture} in {Collagen} {Mechanics}},
	volume = {114},
	issn = {0006-3495, 1542-0086},
	url = {https://www.cell.com/biophysj/abstract/S0006-3495(18)30538-1},
	doi = {10.1016/j.bpj.2018.04.043},
	language = {English},
	number = {11},
	urldate = {2025-06-13},
	journal = {Biophysical Journal},
	author = {Jansen, Karin A. and Licup, Albert J. and Sharma, Abhinav and Rens, Robbie and MacKintosh, Fred C. and Koenderink, Gijsje H.},
	month = jun,
	year = {2018},
	pmid = {29874616},
	pages = {2665--2678},
}

@article{kumar_ramplot_2025,
	title = {{RamPlot}: a webserver to draw {2D}, {3D} and assorted {Ramachandran} maps},
	volume = {58},
	issn = {1600-5767},
	shorttitle = {{RamPlot}},
	url = {https://journals.iucr.org/j/issues/2025/02/00/jo5117/},
	doi = {10.1107/S1600576725001669},
	language = {en},
	number = {2},
	urldate = {2025-07-07},
	journal = {Journal of Applied Crystallography},
	author = {Kumar, M. and Rathore, R. S.},
	month = apr,
	year = {2025},
	note = {Publisher: International Union of Crystallography},
	pages = {630--636},
}

@article{revell_collagen_2021,
	title = {Collagen fibril assembly: {New} approaches to unanswered questions},
	volume = {12},
	issn = {2590-0285},
	shorttitle = {Collagen fibril assembly},
	url = {https://www.sciencedirect.com/science/article/pii/S2590028521000235},
	doi = {10.1016/j.mbplus.2021.100079},
	urldate = {2025-09-10},
	journal = {Matrix Biology Plus},
	author = {Revell, Christopher K. and Jensen, Oliver E. and Shearer, Tom and Lu, Yinhui and Holmes, David F. and Kadler, Karl E.},
	month = dec,
	year = {2021},
	pages = {100079},
}

@article{paten_flow-induced_2016,
	title = {Flow-{Induced} {Crystallization} of {Collagen}: {A} {Potentially} {Critical} {Mechanism} in {Early} {Tissue} {Formation}},
	volume = {10},
	issn = {1936-0851, 1936-086X},
	shorttitle = {Flow-{Induced} {Crystallization} of {Collagen}},
	url = {https://pubs.acs.org/doi/10.1021/acsnano.5b07756},
	doi = {10.1021/acsnano.5b07756},
	language = {en},
	number = {5},
	urldate = {2025-09-10},
	journal = {ACS Nano},
	author = {Paten, Jeffrey A. and Siadat, Seyed Mohammad and Susilo, Monica E. and Ismail, Ebraheim N. and Stoner, Jayson L. and Rothstein, Jonathan P. and Ruberti, Jeffrey W.},
	month = may,
	year = {2016},
	pages = {5027--5040},
}

@article{canty_actin_2006,
	title = {Actin {Filaments} {Are} {Required} for {Fibripositor}-mediated {Collagen} {Fibril} {Alignment} in {Tendon} *},
	volume = {281},
	issn = {0021-9258, 1083-351X},
	url = {https://www.jbc.org/article/S0021-9258(20)71849-1/abstract},
	doi = {10.1074/jbc.M607581200},
	language = {English},
	number = {50},
	urldate = {2025-09-10},
	journal = {Journal of Biological Chemistry},
	author = {Canty, Elizabeth G. and Starborg, Tobias and Lu, Yinhui and Humphries, Sally M. and Holmes, David F. and Meadows, Roger S. and Huffman, Adam and O'Toole, Eileen T. and Kadler, Karl E.},
	month = dec,
	year = {2006},
	pages = {38592--38598},
}

@article{shoulders_collagen_2009,
	title = {Collagen {Structure} and {Stability}},
	volume = {78},
	issn = {0066-4154, 1545-4509},
	url = {https://www.annualreviews.org/content/journals/10.1146/annurev.biochem.77.032207.120833},
	doi = {10.1146/annurev.biochem.77.032207.120833},
	language = {en},
	number = {Volume 78, 2009},
	urldate = {2025-09-10},
	journal = {Annual Review of Biochemistry},
	author = {Shoulders, Matthew D. and Raines, Ronald T.},
	month = jul,
	year = {2009},
	pages = {929--958},
}

@article{hulmes_radial_1995,
	title = {Radial packing, order, and disorder in collagen fibrils},
	volume = {68},
	issn = {0006-3495},
	doi = {10.1016/S0006-3495(95)80391-7},
	language = {eng},
	number = {5},
	journal = {Biophysical Journal},
	author = {Hulmes, D. J. and Wess, T. J. and Prockop, D. J. and Fratzl, P.},
	month = may,
	year = {1995},
	pmid = {7612808},
	pmcid = {PMC1282067},
	pages = {1661--1670},
}

@article{north_structural_1954,
	title = {Structural {Units} in {Collagen} {Fibrils}},
	volume = {174},
	copyright = {1954 Springer Nature Limited},
	issn = {1476-4687},
	url = {https://www.nature.com/articles/1741142a0},
	doi = {10.1038/1741142a0},
	language = {en},
	number = {4442},
	urldate = {2025-09-10},
	journal = {Nature},
	author = {North, A. C. T. and Cowan, P. M. and Randall, J. T.},
	month = dec,
	year = {1954},
	pages = {1142--1143},
}

@article{veis_limiting_1967,
	title = {A {Limiting} {Microfibril} {Model} for the {Three}-dimensional {Arrangement} within {Collagen} {Fibres}},
	volume = {215},
	copyright = {1967 Springer Nature Limited},
	issn = {1476-4687},
	url = {https://www.nature.com/articles/215931a0},
	doi = {10.1038/215931a0},
	language = {en},
	number = {5104},
	urldate = {2025-09-10},
	journal = {Nature},
	author = {Veis, Arthur and Anesey, Joan and Mussell, Shirley},
	month = aug,
	year = {1967},
	pages = {931--934},
}

@article{smith_molecular_1968,
	title = {Molecular {Pattern} in {Native} {Collagen}},
	volume = {219},
	copyright = {1968 Springer Nature Limited},
	issn = {1476-4687},
	url = {https://www.nature.com/articles/219157a0},
	doi = {10.1038/219157a0},
	language = {en},
	number = {5150},
	urldate = {2025-09-10},
	journal = {Nature},
	author = {Smith, J. W.},
	month = jul,
	year = {1968},
	pages = {157--158},
}

@article{hulmes_quasi-hexagonal_1979,
	title = {Quasi-hexagonal molecular packing in collagen fibrils},
	volume = {282},
	copyright = {1979 Springer Nature Limited},
	issn = {1476-4687},
	url = {https://www.nature.com/articles/282878a0},
	doi = {10.1038/282878a0},
	language = {en},
	number = {5741},
	urldate = {2025-09-10},
	journal = {Nature},
	author = {Hulmes, David J. S. and Miller, Andrew},
	month = dec,
	year = {1979},
	pages = {878--880},
}

@article{huang_charmm36m_2017,
	title = {{CHARMM36m}: an improved force field for folded and intrinsically disordered proteins},
	volume = {14},
	copyright = {2016 Springer Nature America, Inc.},
	issn = {1548-7105},
	shorttitle = {{CHARMM36m}},
	url = {https://www.nature.com/articles/nmeth.4067},
	doi = {10.1038/nmeth.4067},
	language = {en},
	number = {1},
	urldate = {2025-09-11},
	journal = {Nature Methods},
	author = {Huang, Jing and Rauscher, Sarah and Nawrocki, Grzegorz and Ran, Ting and Feig, Michael and de Groot, Bert L. and Grubmüller, Helmut and MacKerell, Alexander D.},
	month = jan,
	year = {2017},
	pages = {71--73},
}

@article{boonstra_charmm_2016,
    author = {Boonstra, Sander and Onck, Patrick R. and van der Giessen, Erik},
    title = {CHARMM TIP3P Water Model Suppresses Peptide Folding by Solvating the Unfolded State},
    journal = {The Journal of Physical Chemistry B},
    volume = {120},
    number = {15},
    pages = {3692-3698},
    year = {2016},
    doi = {10.1021/acs.jpcb.6b01316},
}

@article{shen_protein_2021,
	title = {From {Protein} {Building} {Blocks} to {Functional} {Materials}},
	volume = {15},
	issn = {1936-0851},
	url = {https://doi.org/10.1021/acsnano.0c08510},
	doi = {10.1021/acsnano.0c08510},
	number = {4},
	urldate = {2025-11-17},
	journal = {ACS Nano},
	author = {Shen, Yi and Levin, Aviad and Kamada, Ayaka and Toprakcioglu, Zenon and Rodriguez-Garcia, Marc and Xu, Yufan and Knowles, Tuomas P. J.},
	month = apr,
	year = {2021},
	pages = {5819--5837},
}

@article{gisbert_high-speed_2021,
	title = {High-{Speed} {Nanomechanical} {Mapping} of the {Early} {Stages} of {Collagen} {Growth} by {Bimodal} {Force} {Microscopy}},
	volume = {15},
	issn = {1936-0851},
	url = {https://doi.org/10.1021/acsnano.0c10159},
	doi = {10.1021/acsnano.0c10159},
	number = {1},
	urldate = {2025-03-23},
	journal = {ACS Nano},
	author = {Gisbert, Victor G. and Benaglia, Simone and Uhlig, Manuel R. and Proksch, Roger and Garcia, Ricardo},
	month = jan,
	year = {2021},
	pages = {1850--1857},
}

@article{inamdar_secret_2017,
	title = {The {Secret} {Life} of {Collagen}: {Temporal} {Changes} in {Nanoscale} {Fibrillar} {Pre}-{Strain} and {Molecular} {Organization} during {Physiological} {Loading} of {Cartilage}},
	volume = {11},
	issn = {1936-0851},
	shorttitle = {The {Secret} {Life} of {Collagen}},
	url = {https://doi.org/10.1021/acsnano.7b00563},
	doi = {10.1021/acsnano.7b00563},
	number = {10},
	urldate = {2025-11-14},
	journal = {ACS Nano},
	author = {Inamdar, Sheetal R. and Knight, David P. and Terrill, Nicholas J. and Karunaratne, Angelo and Cacho-Nerin, Fernando and Knight, Martin M. and Gupta, Himadri S.},
	month = oct,
	year = {2017},
	pages = {9728--9737},
}

@article{robertson_improved_2015,
	title = {Improved {Peptide} and {Protein} {Torsional} {Energetics} with the {OPLS}-{AA} {Force} {Field}},
	volume = {11},
	issn = {1549-9618},
	url = {https://doi.org/10.1021/acs.jctc.5b00356},
	doi = {10.1021/acs.jctc.5b00356},
	number = {7},
	urldate = {2025-11-13},
	journal = {Journal of Chemical Theory and Computation},
	author = {Robertson, Michael J. and Tirado-Rives, Julian and Jorgensen, William L.},
	month = jul,
	year = {2015},
	pages = {3499--3509},
}

@article{thompson_general_2009,
	title = {General formulation of pressure and stress tensor for arbitrary many-body interaction potentials under periodic boundary conditions},
	volume = {131},
	issn = {0021-9606},
	url = {https://doi.org/10.1063/1.3245303},
	doi = {10.1063/1.3245303},
	number = {15},
	urldate = {2025-09-12},
	journal = {The Journal of Chemical Physics},
	author = {Thompson, Aidan P. and Plimpton, Steven J. and Mattson, William},
	month = oct,
	year = {2009},
	pages = {154107},
}

@article{lorenzo_elastic_2005,
	title = {Elastic properties, {Young}'s modulus determination and structural stability of the tropocollagen molecule: a computational study by steered molecular dynamics},
	volume = {38},
	issn = {0021-9290},
	shorttitle = {Elastic properties, {Young}'s modulus determination and structural stability of the tropocollagen molecule},
	url = {https://www.sciencedirect.com/science/article/pii/S0021929004003550},
	doi = {10.1016/j.jbiomech.2004.07.011},
	number = {7},
	urldate = {2025-10-02},
	journal = {Journal of Biomechanics},
	author = {Lorenzo, Alicia Claudia and Caffarena, Ernesto Raúl},
	month = jul,
	year = {2005},
	pages = {1527--1533},
}

@article{sasaki_stress-strain_1996,
	title = {Stress-strain curve and young's modulus of a collagen molecule as determined by the {X}-ray diffraction technique},
	volume = {29},
	issn = {0021-9290},
	url = {https://www.sciencedirect.com/science/article/pii/0021929095001107},
	doi = {10.1016/0021-9290(95)00110-7},
	number = {5},
	urldate = {2025-10-02},
	journal = {Journal of Biomechanics},
	author = {Sasaki, Naoki and Odajima, Singo},
	month = may,
	year = {1996},
	pages = {655--658},
}

@article{cusack_determination_1979,
	title = {Determination of the elastic constants of collagen by {Brillouin} light scattering},
	volume = {135},
	issn = {0022-2836},
	url = {https://www.sciencedirect.com/science/article/pii/0022283679903395},
	doi = {10.1016/0022-2836(79)90339-5},
	number = {1},
	urldate = {2025-10-02},
	journal = {Journal of Molecular Biology},
	author = {Cusack, S. and Miller, A.},
	month = nov,
	year = {1979},
	pages = {39--51},
}

@article{van_der_rijt_micromechanical_2006,
	title = {Micromechanical {Testing} of {Individual} {Collagen} {Fibrils}},
	volume = {6},
	copyright = {Copyright © 2006 WILEY-VCH Verlag GmbH \& Co. KGaA, Weinheim},
	issn = {1616-5195},
	url = {https://onlinelibrary.wiley.com/doi/abs/10.1002/mabi.200600063},
	doi = {10.1002/mabi.200600063},
	language = {en},
	number = {9},
	urldate = {2025-10-02},
	journal = {Macromolecular Bioscience},
	author = {van der Rijt, Joost A. J. and van der Werf, Kees O. and Bennink, Martin L. and Dijkstra, Pieter J. and Feijen, Jan},
	year = {2006},
	pages = {697--702},
}

@article{shen_stress-strain_2008,
	title = {Stress-{Strain} {Experiments} on {Individual} {Collagen} {Fibrils}},
	volume = {95},
	issn = {0006-3495},
	url = {https://www.sciencedirect.com/science/article/pii/S0006349508785345},
	doi = {10.1529/biophysj.107.124602},
	number = {8},
	urldate = {2025-10-02},
	journal = {Biophysical Journal},
	author = {Shen, Zhilei L. and Dodge, Mohammad Reza and Kahn, Harold and Ballarini, Roberto and Eppell, Steven J.},
	month = oct,
	year = {2008},
	pages = {3956--3963},
}

@article{orgel_microfibrillar_2006,
	title = {Microfibrillar structure of type {I} collagen in situ},
	volume = {103},
	url = {https://www.pnas.org/doi/abs/10.1073/pnas.0502718103},
	doi = {10.1073/pnas.0502718103},
	number = {24},
	urldate = {2025-10-16},
	journal = {Proceedings of the National Academy of Sciences},
	author = {Orgel, Joseph P. R. O. and Irving, Thomas C. and Miller, Andrew and Wess, Tim J.},
	month = jun,
	year = {2006},
	pages = {9001--9005},
}

@article{badar_nonlinear_2025,
	title = {Nonlinear {Stress}-{Induced} {Transformations} in {Collagen} {Fibrillar} {Organization}, {Disorder} and {Strain} {Mechanisms} in the {Bone}-{Cartilage} {Unit}},
	volume = {12},
	issn = {2198-3844},
	url = {https://onlinelibrary.wiley.com/doi/abs/10.1002/advs.202407649},
	doi = {10.1002/advs.202407649},
	language = {en},
	number = {1},
	urldate = {2025-10-16},
	journal = {Advanced Science},
	author = {Badar, Waqas and Inamdar, Sheetal R. and Fratzl, Peter and Snow, Tim and Terrill, Nicholas J. and Knight, Martin M. and Gupta, Himadri S.},
	year = {2025},
	pages = {2407649},
}

@article{vassaux_heterogeneous_2024,
	title = {Heterogeneous {Structure} and {Dynamics} of {Water} in a {Hydrated} {Collagen} {Microfibril}},
	volume = {25},
	issn = {1525-7797},
	url = {https://doi.org/10.1021/acs.biomac.4c00183},
	doi = {10.1021/acs.biomac.4c00183},
	number = {8},
	urldate = {2025-10-16},
	journal = {Biomacromolecules},
	author = {Vassaux, Maxime},
	month = aug,
	year = {2024},
	pages = {4809--4818},
}

@article{thompson_lammps_2022,
	title = {{LAMMPS} - a flexible simulation tool for particle-based materials modeling at the atomic, meso, and continuum scales},
	volume = {271},
	issn = {0010-4655},
	url = {https://www.sciencedirect.com/science/article/pii/S0010465521002836},
	doi = {10.1016/j.cpc.2021.108171},
	urldate = {2025-10-16},
	journal = {Computer Physics Communications},
	author = {Thompson, Aidan P. and Aktulga, H. Metin and Berger, Richard and Bolintineanu, Dan S. and Brown, W. Michael and Crozier, Paul S. and in 't Veld, Pieter J. and Kohlmeyer, Axel and Moore, Stan G. and Nguyen, Trung Dac and Shan, Ray and Stevens, Mark J. and Tranchida, Julien and Trott, Christian and Plimpton, Steven J.},
	month = feb,
	year = {2022},
	pages = {108171},
}

@article{thijssen_necking_2023,
	title = {Necking and failure of a particulate gel strand: signatures of yielding on different length scales},
	volume = {19},
	shorttitle = {Necking and failure of a particulate gel strand},
	url = {https://pubs.rsc.org/en/content/articlelanding/2023/sm/d3sm00681f},
	doi = {10.1039/D3SM00681F},
	language = {en},
	number = {38},
	urldate = {2025-10-16},
	journal = {Soft Matter},
	author = {Thijssen, Kristian and B. Liverpool, Tanniemola and Patrick Royall, C. and L. Jack, Robert},
	year = {2023},
	pages = {7412--7428},
}

\end{document}